\documentclass[14pt]{article}
\usepackage{amsmath,amsfonts,amssymb,graphicx,epsfig}
\usepackage{colordvi}
\usepackage[hypertex]{hyperref}
\usepackage{amsbsy}\newcommand{\mbf}{\boldsymbol}

\def\ba{\begin{eqnarray}}
\def\ea{\end{eqnarray}}

\begin{document}


\title{\bf Testing CPT- and Lorentz-odd electrodynamics with waveguides}
\author{ {Andr\'{e} H. Gomes, Jakson M. Fonseca, Winder A. Moura-Melo and Afr\^{a}nio R. Pereira} \\
\it \small \it Departamento de F\'{\i}sica, Universidade Federal de Vi\c{c}osa\\ \small \it 36570-000, Vi\c{c}osa, Minas Gerais, Brazil.}

\date{}

\maketitle

\begin{abstract}
We study CPT- and Lorentz-odd electrodynamics described by the Standard Model Extension. Its radiation is confined to the geometry of a hollow conductor waveguide open along the $z$-axis. In a special class of reference frames, with vanishing both $0$-th and $z$ components of the background field, $(k_{\rm AF})^\mu$,  we determine a number of {\em huge and macroscopically detectable} effects on the confined waves spectra, compared to standard results. Particularly, if  $(k_{\rm AF})^\mu$ points along the $x$ (or $y$) direction, only transverse electric modes, with $E_z=0$, should be observed propagating throughout the guide, while all the transverse magnetic, $B_z=0$, are absent. Such a strong mode suppression makes waveguides quite suitable to probe these symmetry violations using a simple and easily reproducible apparatus.
\end{abstract}

\noindent Corresponding author: Andr\'{e} H. Gomes\\
E-mails: andre.gomes@ufv.br, andrehgomes@gmail.com
\newpage

\section{Introduction and Motivation}

\indent Symmetry is one of the most powerful ideas in the description of natural phenomena. Invariance under rotations is perhaps the commonest symmetry and its relevance in classifying crystal structure is widely recognized. Other examples include space-time translations, yielding energy-momentum conservation, and gauge-invariance, which ensures charge conservation in electrodynamics. In turn, the Standard Model of elementary particles and interactions is also known to be invariant under Lorentz and CPT transformations. Although these latter symmetries have been intensively tested and confirmed by several highly accurate experiments, a number of recent proposals claim that one (or even both) of them is (are) {\em  not exact}; rather, they appear to be violated by {\em extremely small} deviations.

\indent One of the most studied frameworks incorporating these violations is the so-called Standard Model Extension (SME), an effective low-energy action comprised of all the possible deviations from the Standard Model that arise from high-energy string-type theories and that respect the gauge symmetry of the Standard Model, $SU(3)\times SU(2)\times U(1)$, the power-counting renormalizability, and is coordinate-independent. This last requirement implies that the SME is invariant under observer-type Lorentz transformations (but not under particle-like ones, so usual Lorentz invariance no longer holds), besides of being invariant under space-time translations, thus conserving energy and momentum. Additional requirements like causality, unitarity, hermiticity, etc, could eventually be imposed yielding more restrict models \cite{Colladay-Kost-PRD55-6760-1997-PRD58,Colladay-Kost-PRD-58-1998-116002,Kost-Lehnert-PRD63-065008-2001}. The search for Lorentz-violation, in turn, has received considerable attention in the last few years and a number of mechanisms for probing it has been proposed. Among them we may quote those dealing with fermions, especially electrons \cite{matter-electrons1,matter-electrons2} and neutrinos \cite{matter-neutrinos1,matter-neutrinos2,matter-neutrinos3}, while for the gauge sector proposals predict small deviations in Cerenkov \cite{Cerenkov1,Cerenkov2,CerenkovCPTeven1,CerenkovCPTeven2,CerenkovCPTeven3,Hohensee} and  synchrotron radiations \cite{syncroton-radiation1,syncroton-radiation2,syncroton-radiation3,syncroton-radiation4}, confined waves in cavities and waveguides \cite{Mewes-Petroff-PRD,Mewes-PRD-78-2008,Tobar-etal-PRD71-025004-2005,tobar.new,eisele}, black-body-like spectra \cite{nossoPLB2009,CMB1,CMB2}, photon-splitting possibility \cite{photonsplitting1,photonsplitting2}, among others \cite{outrasrefs1,outrasrefs2,outrasrefs3,jakson-daniel}. However, despite of several attempts, Lorentz symmetry remains strong and no contrary experimental evidence has appeared so far. Actually, this symmetry-breaking extends to a broader scenario. The mechanisms of violation should certainly point to a path towards a {\em unified theory}, in which gravity appears consistently accommodated along with the other fundamental interactions.

\indent This work has been motivated by the following question: could such small violations somehow give rise to large and easily detectable effects? Our present investigation that deal with classical radiation, coming from Lorentz- and CPT-odd electrodynamics within the SME and confined to the geometry of a hollow conductor waveguide, provides an affirmative answer. In this situation, the very small violating parameter, a constant vector, $(k_{\rm AF})_\mu$, remarkably yields {\em huge modifications} to the confined waves spectra, as compared to the usual electrodynamics, at least in reference frames where $(k_{\rm AF})_\mu$ is pure space-like, $(k_{\rm AF})_0=0$. By recalling the absence of such {\em macroscopic} effects in this widely used apparatus, our results inevitably argue against these claimed violations. Otherwise, all waveguide experiments must have been performed in reference frames where $(k_{\rm AF})_0\neq0$. If this latter possibility applies, these violations could be probed by performing a waveguide experiment in such a special frame.

\section{The model and its basic features}

\indent The Abelian pure gauge sector of the SME is described by the action obtained from the Lagrangian below\footnote{ We adopt the Minkowski metric with ${\rm diag}\,(\eta_{\mu\nu})=(+;-,-,-)$; $\mu, \nu, \,{\rm etc.}\,= 0,1,2,3$, while $i,j,\, {\rm etc.}\,=1,2,3$; natural units are also assumed, so that $c=\hbar=1$, etc.. }:
	\begin{equation} \label{action}
	{\cal L}=-\frac14 F_{\mu\nu} F^{\mu\nu} -\frac12 (k_{\rm AF})_\mu A_\nu \tilde{F}^{\mu\nu} -\frac14 (k_{\rm 														F})_{\alpha\beta\mu\nu}F^{\alpha\beta}F^{\mu\nu}.
	\end{equation}
From the Lagrangian above there follow the equations of motion $\partial_\mu F^{\mu\nu}=(k_{\rm AF})_\mu \tilde{F}^{\mu\nu} + (k_{\rm F})^{\nu\alpha\beta\gamma} \partial_\alpha F_{\beta\gamma}$, while the geometrical ones remain unaffected, $\partial_\mu\tilde{F}^{\mu\nu}=0$, with $\tilde{F}^{\mu\nu}=\frac12 \epsilon^{\mu\nu\alpha\beta}F_{\alpha\beta}$ (magnetic sources may be consistently inserted, as done in Ref. \cite{nossoPRD2007}). Now, $(k_{\rm AF})_\mu$ and $(k_{\rm F})_{\alpha\beta\mu\nu}$ are rank-1 and rank-4 tensor-type objects, whose canonical dimensions are $[{\rm mass}]^1$ and $[{\rm mass}]^0$, respectively. Once they are non-dynamical constant quantities, they do not properly transform under (particle-like) space-time transformations, consequently Lorentz symmetry is not respected. In addition, $(k_{\rm AF})_\mu$ brings about a further asymmetry once its term is not CPT-invariant \cite{CFJ}. Usually, we assume that such parameters induce very small Lorentz-odd background effects that are supposed to be reminiscent of the very beginning of the Universe, and presumably described by some string-type model (other proposals include varying couplings \cite{bertolami.lehnert}, non-trivial space-time topology \cite{cpt.anomaly}, non-commutative quantum field theories \cite{non.commutative}, among others). To prevent spurious huge enlarging in such parameters it is further assumed that the model above is physically suitable only in that set of reference frames in which their values are very small (these are the so-called concordant frames, for instance, those moving non-relativistically to the Earth \cite{Kost-Lehnert-PRD63-065008-2001}). For example, $(k_{\rm AF})_\mu$ is currently bounded to be $\lesssim 10^{-43}\, {\rm GeV}$ \cite{new.limit.k.af}, while typical maximum values for $(k_{\rm F})_{\alpha\beta\mu\nu}$ lies around $\lesssim 10^{-31} - 10^{-28}$ \cite{Kost-Mewes-PRL99-011601-2007}. Their effects are expected to be very small, as reported by theoretical results, from both classical and quantum analysis, which are often proportional to powers of these parameters. Of course, large effects coming from small modifications in (linear) theories are not expected and their appearance is rare and counter-intuitive. Indeed, even similar confined waves coming from the non-linear Born-Infeld model are predicted to give only very small deviations from usual results \cite{waveguideBI}. Analogously, only small deviations appear if we consider the CPT-even case, or even the present one in other kinds of waveguides, like coaxial-cables \cite{workinprogress}. Once we are able to show that despite its small size the present violations give rise to huge and readily detectable effects our results become important. To be specific, monochromatic waves confined to a hollow conductor waveguide are such that combined restrictions imposed by the symmetry-breaking along with those coming from the boundary conditions yield a spectrum inside the guide consisting of only a unique set of modes; all the others, observed in the standard electromagnetism, are completely suppressed.

\section{Radiation confined in waveguides}

\indent Firstly, let us recall that in Maxwell electrodynamics a monochromatic wave, with frequency $\omega$, tra\-ve\-ling along a given direction, say $z$, is such that its associated electric and magnetic amplitudes {\em do not depend} on $z$-coordinate:
	\begin{eqnarray} \label{waveforms}
	\mbf{E} (\mbf{x},t) = \mbf{E} (\mbf{x}_{\bot}) \, e^{i (\kappa z - w t)}, \quad
	\mbf{B} (\mbf{x},t) = \mbf{B} (\mbf{x}_{\bot}) \, e^{i (\kappa z - w t)},
	\end{eqnarray}
where the vector $\mbf{x}_{\bot}\equiv(x,y)$ points along the plane transverse to the guide axis. We use $\kappa$ for the wave number instead of $k$ because confined waves generally have different `dispersion relations' from their free-space counterparts; $\kappa$ depends upon other parameters, like $\omega$, and boundary conditions, as well.

\indent By confining these waves to travel inside a hollow conductor guide with rectangular cross sections $a$ and $b$, along $x$ and $y$, respectively, only those frequencies higher than the {\em cutoff frequency}, $\omega_{mn}=\sqrt{\kappa^2_x + \kappa^2_y}$, can eventually propagate along the guide; all the other appear to be {\em evanescent waves}, falling off rapidly. Besides being no longer {\em transverse}, in the usual sense, such waves present discrete modes along the confined dimensions, with $\kappa$ and $\omega$ satisfying:
	\begin{equation}\label{DispRelGuideUsual}
	\kappa=\sqrt{\omega^2 -\kappa^2_x - \kappa^2_y}\,,
	\end{equation}

\noindent where $\kappa_x=m\pi/a$ and $\kappa_y=n\pi/b$, with $m,n=0,+1,+2,\ldots$. The two fundamental types of modes are referred to as {\em transverse electric} (TE, with $E_z=0$) and {\em transverse magnetic} (TM, $B_z=0$). Actually, TE and TM modes form together a basis for all the possible modes propagating along the guide (if both field components vanish, $E_z=B_z=0$, no wave propagates inside this guide; such transverse electric-magnetic (TEM) modes do appear, and are the fundamental ones in other sorts of guides, like coaxial cables). In this standard case, the confinement inside the guide along with the field equations impose boundary conditions (BC's) on the electromagnetic fields that require the vanishing of tangential electric and normal magnetic amplitudes at the guide borders $\mathcal{S}$, explicitly:
	\begin{equation}\label{BoundaryConditions}
	E_z |_{\mathcal{S}} \equiv 0	\qquad	\textrm{and}	\qquad	\frac{\partial B_z}{\partial n} \Big|_{\mathcal{S}} \equiv 0,
	\end{equation}

\noindent with $\mbf{\hat{n}}$ being a unit vector normal to the borders $\mathcal{S}$ everywhere. Since the BC's are distinct for the electric and magnetic fields, the TE and TM modes are generally different. Below, we quote the explicit forms of the (complex) electromagnetic field amplitudes for TE modes (up to $e^{i(\kappa z -\omega t)}$; TM modes are obtained analogously, namely, $E_z(x,y)=E_0 \,\sin(\kappa_x x)\,\sin(\kappa_y y)$):
	\begin{equation}\label{AmplitudeUsual}
	\left. \begin{array}{l}
	\vspace{0.3cm} 
 	B_z=B_0\, \cos(\kappa_x x)\,\cos(\kappa_y y)\,,\\
	\vspace{0.3cm}
 	B_y=+\displaystyle{\frac{\kappa}{\omega}}E_x=- \displaystyle{ \frac{i\,B_0 \,\kappa\, \kappa_y}{\omega^2 -\kappa^2}} \, \cos(\kappa_x 	x)\,\sin(\kappa_y y)\,,\\
	\vspace{0.3cm} 
 	B_x=-\displaystyle{\frac{\kappa}{\omega}}E_y=- \displaystyle{ \frac{i\,B_0\,\kappa \, \kappa_x}{\omega^2 -\kappa^2} }\, \sin(\kappa_x 	x)\,\cos(\kappa_y y)\,.
	\end{array} \right\}
	\end{equation}

\noindent For example, if $a<b$, then the lowest cutoff frequency occurs for $m=0\,, n=1$, that is, $\omega_{01}=\pi/b$, with all the smaller frequencies ruled out from this guide (for further details, see \cite{Landau-textbook,jackson-textbook}).

\indent In order to realize that such results, namely those concerning TE and TM modes, are profoundly modified whenever $(k_{\rm AF})_\mu$ is non-vanishing, let us rewrite the analogues of Maxwell equations, obtained from Lagrangian (\ref{action}) with $(k_{\rm F})^{\alpha\beta\mu\nu}=0$ (to simplify the notation we adopt $(k_{\rm AF})^\mu\equiv \xi^\mu$ hereafter): 
	\begin{equation} \label{eqsmotion}
 		\mbf{\nabla}\cdot \mbf{E}= -\mbf{\xi}\cdot \mbf{B}, \quad
	 	\mbf{\nabla}\times \mbf{B} -\partial_t \mbf{E}= -\xi_0 \mbf{B}+ \mbf{\xi}\times\mbf{E} ,
	\end{equation}
	\vskip -.5cm
	\begin{equation}\label{eqsgeom}
		\mbf{\nabla}\cdot\mbf{B}=0, \qquad 
	 	\mbf{\nabla}\times\mbf{E}+\partial_t \mbf{B}=\mbf{0}\,.
	\end{equation}

\noindent These equations can be set in more convenient forms, by separating the field components parallel and perpendicular to the guide axis, like below:
	\begin{equation}\label{div.e.new}
		\mbf{\nabla}_{\bot} \cdot \mbf{E}_{\bot} = - \partial_z E_z - (\mbf{\xi}_{\bot} \cdot \mbf{B}_{\bot} + \xi_z B_z),
	\end{equation}
	\begin{equation}\label{rot.b.new}
	\left.\begin{array}{l}
		(i)\,\,\quad	\mbf{\hat{z}} \cdot (\mbf{\nabla}_{\bot} \times \mbf{B}_{\bot}) = \partial_t E_z - \xi_o B_z + \mbf{\hat{z}} \cdot 					(\mbf{\xi}_{\bot} \times \mbf{E}_{\bot}),	\medskip	\\
		(ii)\quad	\partial_z \mbf{B}_{\bot} + \mbf{\hat{z}} \times \partial_t \mbf{E}_{\bot} = \mbf{\nabla}_{\bot} B_z + \xi_o\, 								\mbf{\hat{z}} \times \mbf{B}_{\bot} - \mbf{\xi}_{\bot} E_z + \xi_z \mbf{E}_{\bot},
	\end{array} \right\}	\medskip
	\end{equation}
	\begin{equation}\label{div.b.new}
		\mbf{\nabla}_{\bot} \cdot \mbf{B}_{\bot} = - \partial_z B_z,	\medskip
	\end{equation}
	\begin{equation}\label{rot.e.new}
	\left.\begin{array}{l}
		(i)\,\,\quad	\mbf{\hat{z}} \cdot (\mbf{\nabla}_{\bot} \times \mbf{E}_{\bot}) = -\partial_t B_z,	\medskip	\\
		(ii)\quad	\partial_z \mbf{E}_{\bot} - \mbf{\hat{z}} \times \partial_t \mbf{B}_{\bot} = \mbf{\nabla}_{\bot} E_z,	\medskip
	\end{array} \right\}
	\end{equation}
\noindent where $\mbf{\nabla}_{\bot}\equiv\mbf{\nabla}-\hat{\mbf{z}}\partial_z$ is the transverse $\mbf{\nabla}$-operator and $\mbf{\xi}_{\bot}\equiv(\xi_x,\xi_y)$. The advantage of these expressions is that they make clearer we only need, along with suitable BC's, to determine the axial amplitudes, $E_z$ and $B_z$, in order to completely determine the transverse ones, $(E_x,E_y)$ and $(B_x,B_y)$.

\indent To apply a similar analysis to the CPT- and Lorentz-odd framework, we should ensure that expansions (\ref{waveforms}) remain valid. Indeed, if we directly use (\ref{waveforms}) in the equations above we find an inconsistency among themselves, the removal of which demands that $\xi^\mu$ be confined to the spatial plane perpendicular to the guide axis, say, $\xi^\mu=(\xi^0\equiv0; \xi_x,\xi_y, \xi_z\equiv0)$, as we see next. This inconsistency is found whenever we use the wave forms (\ref{waveforms}) for general $\xi^\mu$, $\xi^\mu=(\xi^o;\mbf{\xi})$, along with eqs. (\ref{div.e.new}) and (\ref{rot.b.new}.i), from which we find (after writing the transverse amplitudes in terms of the axial ones, $E_z$ and $B_z$):
	\begin{eqnarray} \label{badneweqEzBz1}
 	\left(	\nabla^2_{\bot} + w^2 - \kappa^2 - \mu^2	\right) E_z = \left[	\frac{i}{w^2 - \kappa^2}\frac{w}{\kappa}(\xi_z w - \xi_o \kappa) \nabla^2_{\bot} - \mbf{\xi}_{\bot} \cdot \mbf{\nabla}_{\bot} + i\, \frac{\xi_z}{\kappa} (w^2 - \kappa^2)	\right] B_z,
	\end{eqnarray}
	\begin{eqnarray} \label{badneweqEzBz2}
 	\left(	\nabla^2_{\bot} + w^2 - \kappa^2 - \mu^2	\right) E_z = \left[	\frac{i}{w^2 - \kappa^2}\frac{\kappa}{w}(\xi_z w - \xi_o \kappa) \nabla^2_{\bot} - \mbf{\xi}_{\bot} \cdot \mbf{\nabla}_{\bot} + i\, \frac{\xi_o}{w} (w^2 - \kappa^2)	\right] B_z.
	\end{eqnarray}
For ensuring the uniqueness of the fields, eqs. above must equal each other. This is achieved provided that we have $\omega\,\xi_z=\kappa\,\xi_0$ or $\xi_0=\xi_z\equiv0$. The first relation is clearly non-physical, once it constrains the wave quantities $\omega$ and $\kappa$ to the anisotropy in such a way that $\xi_0/\xi_z$ appears to be the `wave velocity', which could acquire arbitrarily large or small values. Thus we must take $\xi_0=\xi_z\equiv0$ for correctly describe such confined waves in this framework. The reason for that lies in the following fact: If we adopt usual plane-wave decomposition for the free electromagnetic field $F_{\mu\nu}(x)=\int d^4 k\, {\cal F}^{\mu\nu}(k) e^{i k_\alpha x^\alpha}$, the equations of motion readily yields the dispersion relation  $(k_\mu k^\mu)^2 + (k_\mu k^\mu)(\xi_\nu \xi^\nu) -(k_\mu \xi^\mu)^2=0$. Now, taking a plane-wave traveling along the $z$-axis,  $k^\mu=(\omega; 0;0; k_z=|\mbf{k}|)$, we see that it is generally described by means of (\ref{waveforms}) only if $k_\mu \xi^\mu\equiv0$, yielding $\xi_o=\xi_z\equiv 0$. This leaves us with $\omega^2=\mbf{k}^2$ and $\omega^2= \mbf{k}^2 +\mbf{\xi}^2_\bot$, which describes both a massless and a massive-type mode, respectively. Thus, restricting the anisotropy to the spatial plane perpendicular to the guide axis allows the free propagation through the $z$ and time dimensions to be described by the exponential part of (\ref{waveforms}) and it may generate a contribution of the form $k^2 + \mbf{\xi}^2_\bot(1\pm1)/2$ to $w^2$; therefore, the effects of the anisotropy, confined to the $x$-$y$ plane, on the wave form  (\ref{waveforms}) are contained in their amplitudes and the dispersion relation. It is noteworthy that the restriction on $\xi^\mu$, to be pure space-like, is quite reasonable once only in this case the radiation has been shown to be suitably quantized; in addition, if $\xi^0\neq0$, a number of troubles come about, like the loss of micro-causality or unitarity of the model \cite{Kost-Lehnert-PRD63-065008-2001,KlinkhamerNPB2001}. In order to work with general $\xi^\mu$, we should find wave forms more general then (\ref{waveforms}) to this Lorentz- and CPT-odd electrodynamics.

\indent A (concordant) reference frame where $\xi^\mu= (0;\, \xi_x\,,\xi_y\,,0)$ can be achieved, in principle, from any other by a suitable boost, making $\xi^0$ vanishing, followed by an appropriate spatial rotation of the guide to set $\xi_z=0$. Assuming $\xi^\mu=(0; \xi_x,\xi_y;0)$ and taking relation (\ref{waveforms}) to eqs. (\ref{div.e.new})-(\ref{rot.e.new}), the field amplitudes can be written entirely in terms of $E_z$ and $B_z$:
	\begin{equation}\label{b.trans}
	\mbf{B}_{\bot}(\mbf{x}_{\bot}; \kappa^{\mu}) = \frac{i}{w^2 - \kappa^2} \left(	\kappa\, \mbf{\nabla}_{\bot} B_z + w\, \mbf{\hat{z}} \times 						\mbf{\nabla}_{\bot} E_z - \kappa\, \mbf{\xi}_{\bot} E_z	\right),	\medskip
	\end{equation}
	\begin{equation}\label{e.trans}
	\mbf{E}_{\bot}(\mbf{x}_{\bot}; \kappa^{\mu}) = \frac{i}{w^2 - \kappa^2} \left(	\kappa\, \mbf{\nabla}_{\bot} E_z - w\, \mbf{\hat{z}} \times 						\mbf{\nabla}_{\bot} B_z + w\, \mbf{\hat{z}} \times \mbf{\xi}_{\bot} E_z	\right),	\medskip
	\end{equation}
\noindent whereas $E_z$ and $B_z$ amplitudes appear coupled as follows:
	\begin{eqnarray} \label{neweqEzBz}
 	\left(	\nabla^2_{\bot} + w^2 - \kappa^2 - \mu^2	\right) E_z = - \mbf{\xi}_{\bot} \cdot \mbf{\nabla}_{\bot} B_z,
	\end{eqnarray}
	\vskip -.4cm
	\begin{equation}\label{neweqBzEz}
 	\left(	\nabla^2_{\bot} + w^2 - \kappa^2	\right)B_z = \mbf{\xi}_{\bot} \cdot \mbf{\nabla}_{\bot} E_z.
	\end{equation}
\noindent As long as $\mbf{\xi}_{\bot} \to \mbf{0}$, we identically recover the usual expressions. Notice also the absence of a mass-like gap, $\mu^2\equiv\mbf{\xi}^2_\bot$, in eq. (\ref{neweqBzEz}), as a reminiscent of the distinct ways that electric and magnetic fields experience the symmetry violations. Once the equations above incorporate only small modifications, it would be expected that their solutions would accordingly be only slightly changed whenever compared to their usual counterparts. However, the story is not so simple because such presumed solutions should satisfy the boundary conditions. Indeed, before a deeper analysis of (\ref{neweqEzBz}) and (\ref{neweqBzEz}), it is important to determine the BC's explicitly on the axial amplitudes. Since Bianchi identities remain unaltered, the BC's on \emph{tangential} electric and \emph{normal} magnetic amplitudes at the guide borders coincide with the usual ones, say (recall that the assumption of perfect conductor yields $\partial_t \mbf{B}|_{\mathcal{S}}\equiv 0$; real conductor may be treated in the usual way):
	\begin{equation}\label{BoundaryConditions.new}
	\left.\begin{array}{l l l l l l}
		\big(\mbf{\nabla}\times\mbf{E}+\partial_t \mbf{B}\big)\big|_{\mathcal{S}}=\mbf{0}	&	\Rightarrow	&	\mbf{\hat{n}} \times (\mbf{E}_\|+\mbf{E}_\bot)|_{\mathcal{S}} = \mbf{0}	&	\Rightarrow	&	\mbf{\hat{n}} \times \mbf{E}_\| |_{\mathcal{S}} = \mbf{0},	&	(i)	\medskip	\\
		\big(\mbf{\nabla}\cdot\mbf{B}\big)\big|_{\mathcal{S}}=0	&	\Rightarrow	&	\mbf{\hat{n}} \cdot (\mbf{B}_\|+\mbf{B}_\bot)|_{\mathcal{S}} = 0	&	\Rightarrow	&	\mbf{\hat{n}} \cdot \mbf{B}_\bot |_{\mathcal{S}} = 0,	&	(ii)
	\end{array} \right\}
	\end{equation}
\noindent where, from (\ref{BoundaryConditions.new}.i), there follows immediately the desired condition on $E_z$:
	\begin{equation}\label{bc.ez}
	E_z |_{\mathcal{S}} \equiv 0,
	\end{equation}
\noindent which equals the usual one, as expected. Now, the BC for $B_z$ can be found from the non-homogeneous equations, which are modified by the anisotropy. Indeed, the modified Amper\`{e}-Maxwell law (\ref{rot.b.new}.ii), applied to the walls of the waveguide, along with condition (\ref{BoundaryConditions.new}.ii), yields:
	\begin{equation}\label{bc.bz.full}
	\frac{\partial B_z}{\partial n} \Big|_{\mathcal{S}} = - \xi_o\,\hat{\mbf{n}}\cdot\hat{\mbf{z}}\times\mbf{B}_\bot |_{\mathcal{S}} + \hat{\mbf{n}}\cdot\mbf{\xi}_\bot E_z |_{\mathcal{S}} - \xi_z\, \hat{\mbf{n}}\cdot\mbf{E}_\bot |_{\mathcal{S}}.	\medskip
	\end{equation}
\noindent Now, it is worthy to notice that the $\xi_o$-term vanishes using (\ref{BoundaryConditions.new}.ii) along with the cyclic property of the triple-product; the $\mbf{\xi}_\bot$-term does not contribute by virtue of (\ref{BoundaryConditions.new}.i). Thus, in principle, only the last term, $-\xi_z\, \hat{\mbf{n}}\cdot\mbf{E}_\bot |_{\mathcal{S}}$, modifies the boundary condition on $B_z$, as compared to the usual one (\ref{BoundaryConditions}). Here, our special frame where $\xi_z \equiv 0$ enters, leaving us with:
	\begin{equation}\label{bc.bz}
	\frac{\partial B_z}{\partial n} \Big|_{\mathcal{S}} \equiv 0.	\medskip
	\end{equation}
\noindent It is noteworthy that the condition imposed on the space-time anisotropy of being purely space-like and pointing perpendicular to the guide axis, $\xi^\mu=(0;\xi_x,\xi_y,0)$, ensures the validity of both the plane wave expansions (\ref{waveforms}) and the usual BC's (\ref{bc.ez}) and (\ref{bc.bz}) in this scenario with CPT and Lorentz violation.

\bigskip

\indent Although there is no standard procedure for solving (\ref{neweqEzBz}) and (\ref{neweqBzEz}), we can gain further insight about their solutions by decoupling them at fourth order derivatives, say:
	\begin{equation}\label{uncoupled}
		\left[	(\nabla^2_{\bot} + w^2 - \kappa^2)(\nabla^2_{\bot} + w^2 - \kappa^2 - \mu^{\,2}) + 																									(\mbf{\xi}_{\bot}\cdot\mbf{\nabla}_{\bot})^2
		\right]
		\left(
		\begin{array}{l}
		E_z	\\
		B_z
		\end{array}
		\right) = 0,	\medskip
	\end{equation}
\noindent For a while, let us set the anisotropy only in the $x$-direction, $\xi_y=0$, making (\ref{uncoupled}) an eigenvalue equation (with $w$ and $k$ intertwined), that can be formally solved for the amplitudes by means of finite double-Fourier series, coming from products and sums of $exp(\pm i\kappa_x x)$ and  $exp(\pm i\kappa_y y)$. Therefore, the physical solutions will be a subset of these series which satisfies eqs. (\ref{neweqEzBz})-(\ref{neweqBzEz}) along with BC's (\ref{bc.ez}) and (\ref{bc.bz}). By inspecting the BC's we realize that the unique non-trivial solution for $E_z$ must read like $\sin(\kappa_x x)\,\sin(\kappa_y y)$, while for $B_z$ it goes like $\cos(\kappa_x x)\, \cos(\kappa_y y)$. However, it is an easy task to check that such a pair of solutions does not solve eqs. (\ref{neweqEzBz})-(\ref{neweqBzEz}) identically. Consequently, no wave with \emph{both} $E_z$ and $B_z$ non-vanishing can travel along the rectangular guide if $\xi_x$ is non-zero (an analogous analysis yields the same conclusion for $\xi_y\neq0$). Formally, it was shown that the BC's are equivalent to Dirichlet condition on $E_z$ and Neumann on $B_z$, like the conventional case \cite{Landau-textbook,jackson-textbook}. Since both fields satisfy eq. (\ref{uncoupled}), but different BC's, it happens that their eigenvalue spectra are different, making simultaneous non-trivial solutions for $E_z$ and $B_z$ not possible. This takes place by virtue of the anisotropy, which makes their spectra different from each other and, alternatively, note that those terms in eqs. (\ref{neweqEzBz})-(\ref{neweqBzEz}) like $\mbf{\xi}_{\bot} \cdot \mbf{\nabla}_{\bot} B_z$ and $\mbf{\xi}_{\bot} \cdot \mbf{\nabla}_{\bot} E_z$ inevitably force $B_z$ and $E_z$ to have converse parity behavior under $(x,y)\to (-x,-y)$, causing a mismatching between these equations and BC's, yielding these consequences to the waveguide spectrum.

\indent Now, the only way to find non-trivial solutions is by making the BC's over $E_z$ and $B_z$ compatible. This can be done if we focus on the most basic, TE and TM modes, as follows. First, we set $\xi_y\equiv0$ and consider TE-type modes ($E_z=0$), so that eq. (\ref{neweqBzEz}) recovers its standard form whereas eq. (\ref{neweqEzBz}) reduces to :
	\begin{equation}\label{restrictionEz}
	\xi_x\, \partial_x\, B_z\equiv 0\,,
	\end{equation}
\noindent stating that $B_z$ {\em does not depend} on $x$-coordinate. It should be stressed that the precise value of $\xi_x$ is not important for ensuring this fact as long as it is non-zero! As a consequence, we find that all the amplitude components must be $x$-independent or vanishing identically, as below (up to $e^{i(\kappa z -\omega t)}$):
	\begin{equation} \label{newsolutions}
	\left.
	\begin{array}{l}
 		B_z = B_0 \cos(\kappa_y y)\,, \\
 		B_y = \displaystyle{\frac{\kappa}{\omega}} E_x=-\displaystyle{\frac{i}{\omega^2 -\kappa^2}} B_0 \, \kappa 																	  \kappa_y\,\sin\left(\kappa_y 			 \,y \right)\,,\\
 		B_x = E_y\equiv 0\,, 
 	\end{array}
	\right\}
	\end{equation}
\noindent with ($\kappa_y\equiv n\pi/b, \,\,   n=+1,+2, \ldots$): 
	\begin{equation} \label{newdisprel}
	\kappa=\sqrt{\omega^2-\kappa^2_y}\,,
	\end{equation}
\noindent which are the counterparts of the TE modes (\ref{AmplitudeUsual}) and their dispersion relation (\ref{DispRelGuideUsual}), with $\kappa_x=(m\pi/a)\, \equiv 0$, say $m\equiv0$. Actually, it can be noted that this result coincides exactly with the usual TE$_{0n}$ mode \cite{Landau-textbook,jackson-textbook}.

\indent If we had taken $\xi_x=0$ and $\xi_y\neq0$ the results could be obtained from (\ref{AmplitudeUsual}) by just setting $n=0$; namely, we would get $\kappa=\sqrt{\omega^2 -\kappa^2_x}$ instead of (\ref{newdisprel}). Note also that the allowed modes appear {\em linearly} polarized parallel to $\vec{\xi}$ (circular-type polarizations are not allowed). It should also be noted that these results hold as long as the anisotropy is confined to the $x$ or $y$ axis. Letting \emph{both} $\xi_x$ \emph{and} $\xi_y$ become non-vanishing makes the trivial solution for $E_z$ the only one possible. On the other hand, for TM ($B_z=0$) modes, whatever is the direction of $\vec{\xi}$ on the $x$-$y$ plane, only the trivial solution shows up, as may be readily checked.

\indent Therefore, from the whole standard spectrum composed by a complete set of TE $\oplus$ TM modes, only a much smaller subset (TE$_{0n}$ and TE$_{m0}$ modes) is allowed to propagate inside the guide as long as the space-time anisotropy raised by a pure space-like $(k_{\rm AF})_\mu$ exists.  What makes such non-trivial modes so special is their P-even character, shared with the guide geometry itself, and expressed by its BC's, inside of which only those type of modes can propagate at all \cite{Mewes-Petroff-PRD}. Therefore, our results should be better attributed to the breaking of discrete symmetries, Parity and Time Reversal, brought about by the $(k_{AF})^\mu$-term in Lagrangian (\ref{action}) (rather than to the Lorentz violation itself, although it is brought about by the same parameter). This lies on the fact that, in conventional situations where the guide is filled with air or other dieletrics, with permissivity $\varepsilon$ and permeability $\mu$, different from $\varepsilon_o$ and $\mu_o$ (vacuum case), Lorentz symmetry is certainly broken, once $\varepsilon\mu \neq c^{-2}$, while the observed spectrum is only smoothly distorted from the ideal one. Another way to see the specialty of the non-trivial solutions (\ref{newsolutions}) is by considering the electromagnetic energy-momentum tensor for this framework, which is augmented by the extra term $\Delta \Theta_{\mu\nu}=\xi_\nu \, A^\alpha \tilde{F}_{\alpha\mu}$. It does vanish for P-even modes like (\ref{newsolutions}), showing that they carry no extra energy-momentum than the usual ones. Considering this, and now returning to eqs. (\ref{b.trans})-(\ref{e.trans}) and (\ref{neweqEzBz})-(\ref{neweqBzEz}), or, equivalently, to the eqs. of motion (\ref{eqsmotion}), we can find the reason why the usual TE$_{0n}$ and TE$_{m0}$ modes can propagate inside the waveguide: These are the unique modes that completely decouple the electromagnetic field and the background vector $\xi^{\mu}$. These facts do not mean that we are not dealing with the space-time anisotropy, \emph{once it still remains} along $x$ or $y$; rather, we have found that such symmetry violations deeply suppress the spectrum inside the guide, making them a very suitable apparatus to probe for such a space-time anisotropy in a special reference frame.

\indent Before closing we should remark that our analysis and results remain valid for realistic situations where the waveguide is made from metals with finite conductivity and/or with imperfections along its walls. Although the details are too lengthy to be presented here, it suffices to recall that finite conductivity only implies a skin depth for the transverse electric and normal magnetic fields yielding a power loss due to the surface-type current (additionally, real good conductors effectively behave as ideal ones for practical purposes). Small imperfections only require a shift in $\kappa_x$ and/or $\kappa_y$ to incorporate them along the walls \cite{Landau-textbook,jackson-textbook}. None of them changes the P-even character of the modes allowed to propagate inside the guide, nor yields suppression of modes, as found here. In addition, TE and/or TM mode suppression in Maxwell usual electrodynamics remains a challenge requiring specific synthetic (meta)materials with negative permittivity and/or permeability \cite{modesuppression1,modesuppression2,modesuppression3,modesuppression4}; but even in these situations only some specific modes, say, some values of $(m, n)$, are ruled out from the whole spectrum. On the other hand, some apparatus allowing P- and T-odd modes have appeared recently, consisting of nanostructured arrays of gammadions, where polarized light interaction resembles light scattering by anyonic matter \cite{PToddmode}.\\

\section{Concluding Remarks}

\indent In summary, we have considered the radiation sector of the SME with both Lorentz and CPT violations. We have shown that, in reference frames where its associated parameter is pure space-like, the behavior of confined monochromatic waves inside a hollow conducting waveguide is such that the violating parameter yields, despite its smallness, a number of {\em macroscopically detectable} changes: From the whole standard spectrum, only a small subset of TE-type modes survives, all the other are completely suppressed in this framework. Since such predicted effects have not been observed, despite the widely usage of waveguide apparatus, then: i) these violations do not concern, at least as dictated by SME (e.g., parametrized by a constant `4-vector'), or; ii) if it exists in nature, as SME considers, then we should search for a special class of reference frames where $(k_{\rm AF})_\mu$ is pure space-like; in such preferred frames, our findings provide a definite way to probe this symmetry-breaking.

\begin{flushleft}
\Large{Acknowledgments}
\end{flushleft}
\vskip .3cm
\indent The authors are grateful to B. Altschul, H. Belich, R. Casana, D.H.T. Franco, J.A. Helay\"el-Neto and D.R. Viana for fruitful discussions, and to H.F. Fumi\~a for computational help. They also thank CAPES, CNPq and FAPEMIG for financial support.

\end{document}